\documentclass[usenatbib,usegraphicx]{mn2e}
\usepackage{amsmath,amssymb}

\title[Post-Newtonian effects of planetary gravity]
{
Post-Newtonian effects of planetary gravity on the perihelion shift
}
\author[
Kei Yamada, Hideki Asada]
{Kei Yamada, Hideki Asada 
\\ 
Faculty of Science and Technology, 
Hirosaki University, Hirosaki 036-8561, Japan}
\begin{document}

\date{Accepted  Received }

\pagerange{\pageref{firstpage}--\pageref{lastpage}} 

\pubyear{2012}

\maketitle

\label{firstpage}

\begin{abstract}
We consider a coplanar system comprised of a massive central body 
(a star), a less massive secondary (a planet) on a circular orbit, 
and a test particle on a bound orbit exterior to that of 
the secondary. The gravitational pull exerted on the test particle 
by the secondary acts as a small perturbation, 
wherefore the trajectory of the particle can be described 
as an ellipse of a precessing perihelion. 
While the apsidal motion is defined overwhelmingly by the Newtonian 
portion of the secondary's gravity, the post-Newtonian portion, too, 
brings its tiny input. We explore whether this input may be of 
any astrophysical relevance in the next few decades. 
We demonstrate that the overall post-Newtonian input of 
the secondary's gravity can be split into two parts. 
One can be expressed via the orbital angular momentum of the secondary, 
another via its orbital radius. Despite some moderately large 
numerical factors showing up in the expressions for these two parts, 
the resulting post-Newtonian contributions from the secondary's
gravity into the apsidal motion of the test particle 
turn out to be small enough to be neglected 
in the near-future measurements. 
\end{abstract}


\section{Introduction}
The perihelion shift of Mercury 
gave an experimental evidence for the theory of general
relativity a century ago, e.g. \citep{Will}. 
This relativistic effect is caused by the solar gravity, 
which is $10^{-8}$ smaller than the Newtonian ones. 
Substantial improvements including Cassini mission have 
been brought into the ephemerides, e.g. \citep{Pitjeva, Pitjeva2}, 
where the post-Newtonian effects in the solar system have been 
considered numerically in a combined manner, e.g. \citep{MD, Simon}. 

We should note that the post-Newtonian effects of 
planetary gravity on a light-like trajectory 
are of astrophysical relevance. 
Two decades ago, the bending angle of light 
by a planetary mass was observed by VLBI \citep{TL}. 
It is interesting to investigate 
the post-Newtonian effects of 
planetary gravity on a particle motion. 
 
We may insist, before doing explicit calculations, 
that even the post-Newtonian effects of the jovian gravity 
should be $10^{-3}$ smaller than 
the solar ones and thus they should be negligible. 
This argument seems reasonable, 
unless a large numerical factor comes in. 
For instance, $10^{-3}$ multiplied by say $6\pi^2$ is 
nearly five percents. 
It is not the best idea to use numerical simulations 
for the purpose of discussing the smaller post-Newtonian effects of 
planetary gravity separately from other major effects. 

The main purpose of this paper is to clarify 
whether a large numerical factor 
is involved in the post-Newtonian effects of 
planetary gravity on the perihelion shift.  
For this purpose, we propose a model calculation 
that is easily accessible. 

This paper is organized as follows. 
In section II, analytical calculations are carried out 
to study the post-Newtonian effects of 
planetary gravity on the perihelion shift. 
The effects are evaluated for the solar planets in section III. 
Section IV is devoted to the conclusion. 
We provide some calculations in Appendices.

Latin indices run from 1 to 3 and 
Greek ones run from 0 to 3. 
Einstein summation convention is used for both of the indices. 
We take the units of $G=c=1$.

\section{Post-Newtonian effects of planetary gravity} 
\subsection{Effective metric by an orbiting planet} 
In the post-Newtonian approximation, 
the line element for $N$ masses is expressed  
in an inertial frame 
as \citep{LL, MTW}  
\begin{align}
ds^2 &= g_{\mu\nu} (t, \boldsymbol{r}) dx^{\mu}dx^{\nu} 
\notag\\
&= 
\biggl[
- 1 + 2\sum_A \frac{m_A}{r_A} 
- 2 \biggl(\sum_A \frac{m_A}{r_A} \biggr)^2
+ 3 \sum_A \frac{m_Av_A^2}{r_A}
\notag\\
&~~~
- 2 \sum_A \sum_{B \neq A} \frac{m_Am_B}{r_AR_{AB}}
\biggr] dt^2 
\notag\\
&~~~ 
+ 2 \times \biggl[
- \sum_A \frac{m_A}{r_A} 
\biggl\{
\frac72 v_{Aj} 
+ \frac12 \frac{(\boldsymbol{v}_A \cdot \boldsymbol{r}_A) r_{Aj} }{r_A^2}
\biggr\}
\biggr] dtdx^j   \notag\\
&~~~
+ \biggl[
1 + 2\sum_A \frac{m_A}{r_A} 
\biggr] \delta_{ij}dx^{i}dx^{j} ,
\label{ds2}
\end{align}
where $(x^{\mu}) = (t, \boldsymbol{r})$ for $\mu=0, 1, 2, 3$,  
the position of each mass $m_A$ is denoted as $\boldsymbol{R}_A$, 
the relative vectors and distances are defined as 
$\boldsymbol{r}_A \equiv \boldsymbol{r} - \boldsymbol{R}_A$, 
$\boldsymbol{R}_{AB} \equiv \boldsymbol{R}_A - \boldsymbol{R}_B$, 
$r_A \equiv |\boldsymbol{r}_A|$, 
$R_{AB} \equiv |\boldsymbol{R}_{AB}|$ 
and $r \equiv |\boldsymbol{r}|$ 
and the velocity is $\boldsymbol{v}_A$.

The terms in front of $dt dx^j$ in the R.H.S. of Eq. (\ref{ds2}) 
make a contribution as shown in the Appendix A. 
It is caused by the orbital angular momentum of 
both the central object and the secondary one. 
Under the above approximations, however, 
the orbital angular momentum of the central object 
is much smaller than that of the secondary one, 
so that it can be neglected (See Appendix B). 
This effect is distinct from the Lense-Thirring effect 
that comes from a single spinning object \citep{LT}.

\begin{figure}
\includegraphics[width=8cm]{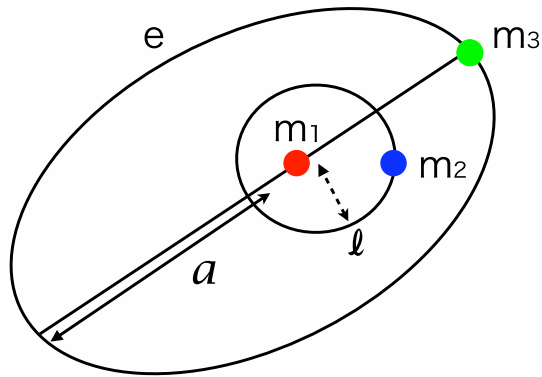}
\caption{ 
Schematic figure of the three masses $m_1$, $m_2$ and $m_3$. 
The mean orbital radius of the secondary mass 
is denoted as $\ell$. 
The semi-major axis and the eccentricity of the Keplerian orbit 
for the third body are denoted as $a$ and $e$. 
}
\label{f1}
\end{figure}

We consider three masses $m_A$ ($A=1, 2, 3$). 
See Figure $\ref{f1}$ for the configuration of the system and our notation. 
For simplicity, approximations are made for the system: \\
\noindent
(i) $m_1 \gg m_2 \gg m_3$, \\
\noindent
(ii) $\ell \ll a$ ($R_2 \ll r$ at the third body position), \\
\noindent
(iii) on the same orbital plane (coplanar), \\
\noindent
(iv) a circular motion of the secondary mass, \\
\noindent
where the orbital radius of the secondary object is denoted as $\ell$
and 
the semi-major axis of the third mass orbit is denoted as $a$. 
The approximation (i) implies that 
the third body can be treated as a test particle 
in the spacetime produced by the two-body system 
of $m_1$ and $m_2$. 
The approximation (ii) suggests that the orbital period 
of the third body is much longer than that of the secondary one 
according to Kepler's third law. 
The approximation (iv) means that the primary mass 
is also moving on circular orbit around 
the common center of mass, 
because $m_3$ is a test mass. 

The post-Newtonian center of mass may affect 
the multipole expansion. 
Fortunately, the correction by the post-Newtonian center 
of mass vanishes for the circular motion \citep{LW}.

The slowly-moving body feels the gravitational field 
that is averaged over fast-changing fluctuations 
caused by the moving primary and secondary objects. 
For discussing the motion of the third body, therefore, 
it is convenient to use the time averaging 
$< \quad >_t \equiv P^{-1} \int dt$ 
for the orbital period $P$. 
For a circular motion, 
this averaging is the same as the angular one 
$< \quad >_{\vartheta} \equiv (2\pi)^{-1} \int d\vartheta$. 
This simplifies the following calculations. 
The averaging is denoted simply as  $< \quad >$. 

Under the approximations (i)-(iv) with the averaging, 
the averaged metric becomes 
\begin{align}
<ds^2> 
&=
\biggl[
- 1 + 2\frac{m_{tot}}{r} \biggl(1 - \frac12\frac{m_2\ell^2}{\ell^3}\biggr)
+ \frac12 \frac{m_2\ell^2}{r^3}
\notag\\
&~~~
- 2\frac{m_{tot}^2}{r^2} - \frac{m_{tot}m_2\ell^2}{r^4} 
\biggr] dt^2 
\notag\\
&~~~
+ \biggl[
1 + 2\frac{m_{tot}}{r} + \frac12 \frac{m_2\ell^2}{r^3}
\biggr] \delta_{ij}dx^idx^j . 
\label{1PN-met}
\end{align}
The calculations for deriving this are given in the Appendix C, 
where our estimate shows that the post-Newtonian effects 
stemming from the circular motion of the primary can be neglected.

Let us introduce the polar coordinates as 
$dr^2 + r^2 (d\theta^2 + \sin^2\theta d\varphi^2) 
= \delta_{ij}dx^idx^j$. 
For a radial coordinate $R$ to be the circumference radius, 
we introduce the isotropic coordinates as 
\begin{align}
R^2 \equiv 
\biggl( 1 + 2\frac{m_{tot}}{r} + \frac12 \frac{m_2\ell^2}{r^3} \biggr)
r^2 . 
\end{align}

By using this transformation, we reach  
the effective metric acting on the third mass as 
\begin{align}
<ds^2> &= 
\biggl(
- 1 + \frac{2m_{tot}}{R} + \frac{m_2\ell^2}{2R^3}
+\frac{m_{tot}m_2\ell^2}{R^4} 
- \frac{m_{tot}m_2\ell^2}{\ell^3R}
\biggr) dt^2 
\notag\\
&~~~
+ \biggl( 1 + \frac{2m_{tot}}{R} + \frac{3m_2\ell^2}{2R^3} \biggr) dR^2
+ R^2 d\varphi^2 \notag\\
&= 
\biggl[
- 1 + \frac{r_s}{R} \biggl(1 - \frac{Q}{\ell^3} \biggr)
+ \frac{Q}{R^3} +\frac{r_sQ}{R^4}
\biggr] dt^2 
\notag\\
&~~~
+ \biggl( 1 + \frac{r_s}{R} + \frac{3Q}{R^3} \biggr) dR^2
+ R^2 d\varphi^2 ,
\label{ds^2}
\end{align}
where we have used $\theta = \pi/2 = \mbox{const.}$ 
for the coplanar configuration. 
Here, we defined the effective Schwarzschild radius as 
$r_S = 2 m_{tot}$ and 
the effective moment induced by the secondary body as 
\begin{equation}
Q \equiv \frac{m_2 \ell^2}{2} . 
\label{Q}
\end{equation}
It follows that the line element given by Eq. (\ref{ds^2}) 
agrees with the Schwarzschild metric on the equatorial plane 
if $Q=0$.

\subsection{Motion in the effective potential} 
Equation (\ref{ds^2}) gives the Lagrangian for the third body as  
\begin{align}
L =& 
- \biggl[
1 - \frac{r_s}{R} \biggl(1 - \frac{Q}{\ell^3} \biggr)
- \frac{Q}{R^3} - \frac{r_sQ}{R^4}
\biggr] \dot{t}^2 
\notag\\
&
+ \biggl( 1 + \frac{r_s}{R} + \frac{3Q}{R^3} \biggr) \dot{R}^2
+ R^2 \dot{\varphi}^2 ,
\label{1PNL-Q}
\end{align}
where the dot denotes the derivative with respect to 
the proper time of the test mass as $d/d\tau$. 

We get constants of motion for this Lagrangian as 
\begin{align}
\varepsilon &\equiv
\biggl[
 1 - \frac{r_s}{R} \biggl(1 - \frac{Q}{\ell^3} \biggr)
- \frac{Q}{R^3} - \frac{r_sQ}{R^4} \biggr] \dot{t} , \\
j &\equiv R^2 \dot{\varphi} .
\label{jj}
\end{align}
Substituting these constants into Eq. (\ref{1PNL-Q}) leads to the
energy integral as 
\begin{align}
&\biggl(1 + \frac{2Q}{R^3}\biggr) \dot{R}^2 
- \frac{r_s}{R} \biggl(1 - \frac{Q}{\ell^3} \biggr)
- \frac{Q}{R^3} - \frac{r_sQ}{R^4}
\notag\\
&
+ \frac{j^2}{R^2}\biggl(1 - \frac{r_s}{R} - \frac{Q}{R^3} \biggr)
= \varepsilon^2 - 1 . 
\label{1PN-E}
\end{align}
Note that this includes both the Newtonian effect with a quadrupole
potential (expressed as $-Q/R^3$ at the third term of
Eq. (\ref{1PN-E})) and 
the Schwarzschild case at the first post-Newtonian order.

It is convenient to introduce 
\begin{equation}
u \equiv \frac1R . 
\end{equation}
Using $dR/d\tau = (dR/d\varphi) (d\varphi/d\tau)$ 
and Eq. (\ref{jj}) for Eq. (\ref{1PN-E}), 
we formally obtain 
\begin{align}
\biggl(\frac{du}{d\varphi} \biggr)^2 
= F(u) , 
\label{F(u)}
\end{align}
where $F(u)$ is a function of $u$ with coefficients 
including $r_s$ and $Q$.

\subsection{Relativistic perihelion shift at $O(Q)$} 
Let $r_a$ and $r_p$ denote the aphelion and perihelion 
in the Keplerian orbit of the third mass 
(the semi-major axis $a$ and the eccentricity $e$). 
We define $u_a \equiv r_a^{-1}$ and 
$u_p \equiv r_p^{-1}$, 
which are related with the orbital elements 
in the Newtonian limit as 
\begin{align}
u_a &= \frac{1}{a(1+e)} ,
\label{u_a}\\
u_p &= \frac{1}{a(1-e)} . 
\label{u_p}
\end{align}
 
The perihelion shift for ($r = r_p \to r_a \to r_p$) 
is written as \citep{Weinberg} 
$\Delta\varphi = 2 \int_{r_p}^{r_a}d\varphi - 2\pi$, 
where the factor 2 comes from the fact that 
the averaged metric by Eq. (\ref{ds^2}) is axisymmetric.

By integrating Eq. (\ref{F(u)}), 
the perihelion shift of the third body 
($u_a \to u_p \to u_a$) is expressed at $O(Q)$ as 
\begin{equation}
\Delta \varphi_Q = \left(24 + \frac{171}{2}e^2 + O(e^4)\right) 
\pi\frac{Q}{a^3} . 
\label{Delta-varphi}
\end{equation}
Note that $r \gg R_A$ leads to $e \ll 1$ 
and hence the above expression is valid for 
small eccentricity $e \ll 1$. 
Eq. (\ref{Delta-varphi}) yields the perihelion shift rate as 
\begin{align}
\dot{\varpi}_Q 
&=
\left(24 + \frac{171}{2}e^2 + O(e^4)\right) 
\frac{Q}{a^3} \frac{\pi}{P}
\notag\\
&=
\left(6 + \frac{171}{8}e^2 + O(e^4)\right) 
\frac{m_2 \ell^2}{a^3} n , 
\label{dot-omega}
\end{align}
where $n = 2\pi P^{-1}$ is the mean motion of the third mass. 
The post-Newtonian correction to $P$ could cause 
the 2PN correction to $\dot{\varpi}$. 

\section{Solar planets}
Eqs. (\ref{Delta-varphi}) and (\ref{dot-omega}) involve 
not huge but moderately large numbers such as $24 \pi$ 
in the coefficients. 
Therefore, we make an estimation of the corrections 
for the solar planets. 

The Jupiter has the largest mass and orbital angular momentum 
among the solar planets. 
First, we choose the secondary object as the Jupiter 
and estimate the magnitude of the jovian effects on the Saturn. 
Table \ref{table1} shows the planetary effects 
due to the orbital angular momentum ($\dot{\varpi}_L$) 
and the induced quadruple moment ($\dot{\varpi}_Q$). 
The orbital angular momentum makes a slightly larger 
effect than the induced quadruple moment. 

Next, let us consider inner planets as another example. 
The post-Newtonian effects of the Mercury on the Venus are 
$-1.4 \times 10^{-3}$ and $8.2 \times 10^{-4}$ mas/cy, 
respectively. 
The estimated values are smaller by two (or more) digits 
than the uncertainties in the current ephemerides 
\citep{Pitjeva, Pitjeva2, Pitjeva3}. 

The above corrections to the perihelion shift for thirty years 
amount to $\sim 1.3 \times 10^{-2} \mbox{[mas/cy]} 
\times 30 \mbox{[yr]} \sim 4 \times 10^{-3} \mbox{[mas]}$ 
for the Saturn 
and $\sim 1.4 \times 10^{-3} \mbox{[mas/cy]} 
\times 30 \mbox{[yr]} \sim 4 \times 10^{-4} \mbox{[mas]}$ 
for the Venus.  
Therefore, the post-Newtonian effects of the planets 
on the orbital motion are small enough 
to be negligible even for near future measurements.

\begin{table}
\caption{
$\dot{\varpi}_{L}$ and $\dot{\varpi}_{Q}$ 
by Jupiter and its inner planets on the Saturn's perihelion, 
where Eqs. (\ref{dot-omega}) and (\ref{shift-L2}) are used. 
}
  \begin{center}
    \begin{tabular}{lllll}
\hline
planet & $m_2$ [kg] & $\ell$ [AU] 
& $\dot{\varpi}_{L}$ [mas/cy] 
& $\dot{\varpi}_{Q}$ [mas/cy] \\
\hline 
Mercury & $3.30 \times 10^{23}$ & 0.387 
& $-6.09 \times 10^{-7}$ & $7.50 \times 10^{-9}$ \\
Venus & $4.87 \times 10^{24}$ & 0.723 
& $-1.23 \times 10^{-5}$ & $3.86 \times 10^{-7}$ \\
Earth & $5.98 \times 10^{24}$ & 1.00 
& $-1.77 \times 10^{-5}$ & $9.06 \times 10^{-7}$ \\
Mars & $6.42 \times 10^{23}$ & 1.52 
& $-2.35 \times 10^{-6}$ & $2.26 \times 10^{-7}$ \\
Jupiter & $1.90 \times 10^{27}$ & 5.20 
& $-1.28 \times 10^{-2}$ & $7.80 \times 10^{-3}$ \\
\hline
    \end{tabular}
  \end{center}
\label{table1}
\end{table}

\section{Conclusion}
The post-Newtonian effects of planetary gravity 
do influence the perihelion shift. 
Analytical calculations were carried out to show that 
the post-Newtonian effects of planetary gravity 
originate from the orbital angular momentum and 
the orbital radius of a planet. 
Thereby, we argued that the post-Newtonian effects 
are small enough to be negligible in the solar system 
even for near future measurements, 
though a moderately large numerical factor is involved. 
A study beyond the present approximations is left as future work. 

Finally, let us mention a gauge issue of the present result. 
The metric Eq. (\ref{ds2}) is based on the standard PN 
coordinates. Namely, the gauge freedom is fixed. 
The gauge choice at the post-Newtonian order makes 
a difference of $10^{-8}$ at most in the solar system dynamics, 
e.g. \citep{Brumberg, Klioner}. 
Therefore, we should note that our result (e.g. the numerical
coefficients) is dependent on the coordinates that we used.

\section*{Acknowledgments}
We would be grateful to M. Efroimsky for 
his invaluable comments on the manuscript. 
We would like to thank E. V. Pitjeva, Y. Itoh and H. Arakida 
for useful comments and discussions. 
 

\appendix
\section{Effects by the planetary orbital angular momentum} 
Let $h_{0j}$ denote the coefficient of $dt dx^j$ 
in $ds^2$ expressed by Eq. (\ref{ds2}). 
The field $h_{0j}$ is generated by massive bodies 1 and 2. 
In the weak field approximation, $h_{0j}$ is known 
to take the form as \citep{LL}
\begin{equation}
h_{0j} = \frac{2}{r^2} (\boldsymbol{n} \times \boldsymbol{L})_j ,  
\label{h0j}
\end{equation}
where $\boldsymbol{L}$ denotes the vector for 
the total orbital angular momentum of the central two objects. 
The perihelion shift rate by this metric component reads \citep{LL} 
\begin{equation}
\dot{\varpi}_L = - \frac{4L}{a^3 (1-e)^{3/2}} , 
\label{shift-L}
\end{equation}
where $L$ is the spin angular momentum. 
For our case ($m_1 \gg m_2$), $L$ 
corresponds to the orbital angular momentum as 
$m_2 \ell^2 (m_{tot}/\ell^3)^{1/2}$,  
where the orbital angular momentum of the primary object 
is negligible for $m_1 \gg m_2$. 
Substitution of this into Eq. (\ref{shift-L}) 
leads to the perihelion shift rate as 
\begin{equation}
\dot{\varpi}_L =
- \frac{4 m_2 \ell^{1/2}}{a^{3/2} (1-e^2)^{3/2}} n . 
\label{shift-L2}
\end{equation}

\section{Solar and planerary orbital angular momenta} 
The solar and planetary orbital angular momenta 
$[kg \cdot m^2 s^{-1}]$ 
with respect to the solar system's barycenter are 
$1.8 \times 10^{40}$ (Sun), 
$9.2 \times 10^{38}$ (Mercury), 
$1.8 \times 10^{40}$ (Venus), 
$2.7 \times 10^{40}$ (Earth), 
$3.5 \times 10^{39}$ (Mars), 
$1.9 \times 10^{43}$ (Jupiter), 
$7.8 \times 10^{42}$ (Saturn), 
$1.7 \times 10^{42}$ (Uranus) and 
$2.5 \times 10^{42}$ (Neptune), 
where the solar orbital angular momentum 
is caused predominantly by outer planets such as the Jupiter and Saturn. 
The Sun's orbital angular momentum is much 
smaller than that of the Jupiter (and the Saturn). 

\section{Averaged metric}
Let us explain how to obtain Eq. (\ref{1PN-met}). 
The potentials in the metric expressed by Eq. (\ref{ds2}) 
are expanded,  
because $r \gg R_A$ at the third-body position. 
For instance, 
\begin{align}
\frac1{r_A} &= \frac1{|\boldsymbol{r} - \boldsymbol{R}_A|} \notag\\
&=
\frac1r \biggl[
1 
+ \frac{R_A}{r}\boldsymbol{n} \cdot \boldsymbol{n}_A 
- \frac12 \frac{R_A^2}{r^2} \{
1 - 3(\boldsymbol{n} \cdot \boldsymbol{n}_A)^2 
\}
+ \mathrm{O}\biggl( \frac{R_A}{r} \biggr)^3
\biggr] ,
\label{1/r}
\end{align}
where we define unit vectors as 
$\boldsymbol{n} \equiv r^{-1} \boldsymbol{r}$, 
$\boldsymbol{n}_A \equiv R_A^{-1} \boldsymbol{R}_A$, 
One can use 
$m_1R_1 \boldsymbol{n}_1 = -  m_2R_2 \boldsymbol{n}_2$, 
because the origin of the coordinates is chosen as 
the center of mass. 
One can expand the Newtonian potential as 
\begin{align}
\sum_A \frac{m_A}{r_A} &=
\frac{m_{tot}}{r} \notag\\
&~~~
- \frac12 \biggl[
\frac{m_1}{r^3} \{R_1^2
  - 3R_1^2\cos^2\vartheta \}
+ \frac{m_2}{r^3} \{R_2^2
  - 3R_2^2\cos^2\vartheta \} 
\biggr]  
\notag\\
&~~~
+ O\biggl( \frac{R_A}{r} \biggr)^4 , 
\label{M/r}
\end{align}
where the third body is considered a test mass and hence 
we define the total mass as $m_{tot} \equiv m_1+m_2$ 
and 
$\cos\vartheta \equiv \boldsymbol{n} \cdot \boldsymbol{n}_2$ 
defines the angle $\vartheta$ 
from the direction to the secondary body to the third one 
at the coordinates origin.  
It is clear that the second term in the R.H.S. of Eq. (\ref{M/r}) 
is the quadrupolar part. 
We may think that the post-Newtonian center of mass 
may affect the above expansion. 
Fortunately, the correction by the post-Newtonian center 
of mass vanishes for the circular motion \citep{LW}. 

After being averaged, the Newton-type potential becomes 
\begin{align}
\biggl< \sum_A\frac{M_A}{r_A} \biggr> &=
\frac{m_{tot}}{r} 
+ \frac14 \frac{m_1R_1^2 + m_2R_2^2}{r^3} 
+ \mathrm{O}\biggl( \frac{R_A}{r} \biggr)^4 . 
\label{M/r-2}
\end{align}

The assumption (i) leads to 
\begin{align}
\boldsymbol{R}_1 &= - \frac{m_2}{m_1}\boldsymbol{\ell} 
+ O\left( \frac{m_2^2}{m_1^2} \ell \right) , \\
\boldsymbol{R}_2 &= \boldsymbol{\ell} 
+ O\left( \frac{m_2}{m_1} \ell \right) ,
\end{align}
where $\boldsymbol{\ell} \equiv \boldsymbol{R}_{21}$. 
Hence, we obtain $m_1R_1^2 + m_2R_2^2 = m_2 \ell^2$ 
at the lowest order, 
so that Eq. (\ref{M/r-2}) can be rewritten as 
\begin{align}
\biggl< \sum_A\frac{M_A}{r_A} \biggr> 
\simeq
\frac{m_{tot}}{r} 
+ \frac14 \frac{m_2\ell^2}{r^3} 
+ \mathrm{O}\biggl( \frac{R_A}{r} \biggr)^4 . 
\label{M/r-heikin}
\end{align}

Similarly, under the assumptions (i) and (ii), 
the velocity-dependent part is averaged as 
\begin{align}
\biggl< \sum_A \frac{m_Av_A^2}{r_A} \biggr> 
= \frac{m_{tot}m_2}{\ell r} ,
\end{align}
where for computing this post-Newtonian part 
it is sufficient to make substitutions as 
$v_1^2 = R_1^2 \Omega^2 \to 0$ 
and $v_2^2 = R_2^2 \Omega^2 \to m_{tot} \ell^{-1}$. 
Here, the secondary's angular velocity is denoted as $\Omega$. 
Finally, the second-order part in masses is expanded as 
\begin{align}
\sum_A \sum_{B \neq A} \frac{m_Am_B}{r_AR_{AB}} 
&=
\frac{2m_1m_2}{\ell r} ,
\end{align}
which is not affected by the averaging. 

By combining the above results, therefore, 
the averaged metric is expressed by Eq. (\ref{1PN-met}).

\bsp

\label{lastpage}

\end{document}